# Implications of "peak oil" for atmospheric $CO_2$ and climate

**Pushker A. Kharecha and James E. Hansen**

NASA Goddard Institute for Space Studies and Columbia University Earth Institute, New York, NY 10025, USA

E-mail: pushker@giss.nasa.gov

**Abstract**
Unconstrained $CO_2$ emission from fossil fuel burning has been the dominant cause of observed anthropogenic global warming. The amounts of 'proven' and potential fossil fuel reserves are uncertain and debated. Regardless of the true values, society has flexibility in the degree to which it chooses to exploit these reserves, especially unconventional fossil fuels and those located in extreme or pristine environments. If conventional oil production peaks within the next few decades, it may have a large effect on future atmospheric $CO_2$ and climate change, depending upon subsequent energy choices. Assuming that proven oil and gas reserves do not greatly exceed estimates of the Energy Information Administration, and recent trends are toward lower estimates, we show that it is feasible to keep atmospheric $CO_2$ from exceeding about 450 ppm by 2100, provided that emissions from coal, unconventional fossil fuels, and land use are constrained. Coal-fired power plants without sequestration must be phased out before mid-century to achieve this $CO_2$ limit. It is also important to 'stretch' conventional oil reserves via energy conservation and efficiency, thus averting strong pressures to extract liquid fuels from coal or unconventional fossil fuels while clean technologies are being developed for the era 'beyond fossil fuels'. We argue that a rising price on carbon emissions is needed to discourage conversion of the vast fossil resources into usable reserves, and to keep $CO_2$ beneath the 450 ppm ceiling.

**Keywords:** anthropogenic climate change, carbon dioxide, fossil fuels, global warming, peak oil

# 1. Introduction

M. King Hubbert, the late petroleum geologist and Shell oil company consultant, articulated the notion that oil production would peak when about half of the economically recoverable resource had been exploited. His successful prediction of peak oil production in the continental United States (*Hubbert*, 1956) has encouraged numerous analysts to subsequently apply his model or variations thereof to global oil production. The concept of peak extraction of a finite nonrenewable resource constrained by geology and geography has received support from similar patterns of growth, peak production, and decline of mineral resources (*van der Veen*, 2006), natural gas (*Lam*, 1998), and coal (*Milici and Campbell*, 1997) in specific regions.

There is intense disagreement about when global 'peak oil' might occur, but it is widely accepted that it will occur at some point this century (*Wood et al.*, 2003; *Kerr*, 2005). Despite the obvious relevance of peak oil to future climate change, it has received little attention in projections of future climate change.

In this paper we emphasize the estimated magnitudes of fossil fuel resources (oil, natural gas, and coal), and the relevance of these limitations to the question of how practical it may be to avoid "dangerous anthropogenic interference" with global climate as outlined in the United Nations Framework Convention on Climate Change (*UNFCCC*, 1992). We are motivated by the conclusion of *Hansen et al.* [2007a,b] that "dangerous" climatic consequences are expected at atmospheric $CO_2$ levels exceeding 450 ppm and possibly at even lower levels. Thus we investigate whether atmospheric $CO_2$ can be kept to 450 ppm or less via constraints on the use of coal and unconventional fossil fuel resources.

In estimating atmospheric $CO_2$ levels for given emission scenarios we employ both a linear atmospheric pulse response function (PRF) fit to the Bern carbon cycle model (*Joos et al.*, 1996) and a non-linear mixed-layer PRF fit to the same Bern carbon cycle model. The mixed-layer (or dynamic-sink) PRF allows non-linear ocean carbonate chemistry, which is not included in the atmospheric (static-sink) PRF. The static-sink PRF underestimates $CO_2$ levels for large emission cases, but we include results for this PRF because our main interest in cases that keep atmospheric $CO_2$ at ~450 ppm, this PRF has the advantage of simplicity and transparency, and we have used this function in other studies (Hansen et al., 2007a,b). We also include results for the dynamic-sink PRF.

We do not attempt to resolve the debate about the timing of peak oil or the magnitude of fossil fuel resources; rather, we consider reasonable alternative assumptions. We recognize that the magnitude of recoverable oil and gas resources depends on economic incentives and penalties. Indeed, the disparity between those who believe that fossil fuels are running out and those who foresee large potential reserves most likely relates in part to assumptions based on the 'principle of substitution' (*Marshall*, 1890), the neoclassical economics notion that implies that technology improvements and supply and demand dynamics will allow continual transfer of fossil fuels from resources to reserves.

We suggest that the potential for "dangerous" climatic consequences should influence the degree to which such substitution remains unfettered. Thus we discuss the possible need for a growing price on carbon emissions, if $CO_2$ is to be kept to a low level.

## 2. Methods
### 2.1 Fossil fuel supply estimates and terminology

We use fossil fuel $CO_2$ emissions from the historical (1750-2003) analysis of the United States Department of Energy's Carbon Dioxide Information Analysis Center (CDIAC; *Marland et al.*, 2006). The record is extended through 2005 with data from British Petroleum (*BP*, 2006), with the BP data for each of the three fuels adjusted slightly by a factor near unity such that the BP and CDIAC data coincide exactly in 2003.

Estimates of remaining fossil fuel reservoirs by the United States Energy Information Administration (*EIA*, 2006), World Energy Council (*WEC*, 2007), and the Intergovernmental Panel on Climate Change (*IPCC*, 2001a) are shown in Figure 1. In our calculations we consider combinations of estimates from these groups reflecting a wide range of fossil fuel supplies. All estimates are shown in units of gigatonnes of carbon (1 Gt C = 1 Pg C = $10^{15}$ g C) for the sake of uniformity, as well as in units of atmospheric $CO_2$ equivalent (1 ppm $CO_2 \approx 2.12$ Gt C).

'Proven reserves' are the amounts of these fuels that are estimated to be economically recoverable under current economic and environmental conditions with existing technology. 'Resources' are fossil occurrences whose existence is well-known but whose recoverable magnitudes are less certain; however, they are widely believed to contain immense amounts of fossil energy and carbon (e.g., see *IPCC*, 2001a). 'Reserve growth' is defined by *EIA* [2006] as expected additions to proven reserves (from resources) based on realistic expected improvements in extraction technologies.

Reservoir estimates tend to be larger if they are made under the assumption of very high fuel prices, or if it is assumed that greater technology advances will allow recovery of a much higher percentage of fossil fuels in existing fields. On the other hand, if a substantial carbon price is applied to $CO_2$ emissions in the future, the available reservoir may decrease, as it becomes unprofitable to extract resources from remote locations or to squeeze hard-to-extract resources from existing fields. Because of these uncertainties, we also consider a range of estimates of fossil fuel reservoirs.

'Unconventional' fossil fuels are those that exist in a physical state other than conventional oil, gas and coal. The contribution of unconventional fossil fuels to $CO_2$ emissions is negligible to date (*IPCC*, 2001a). We do not include unconventional fossil fuels in the scenarios that we illustrate, because we are interested primarily in scenarios that cap atmospheric $CO_2$ at 450 ppm or less. However, it should be borne in mind that unconventional fossil fuels could contribute huge amounts of atmospheric $CO_2$, if the world should follow an unconstrained "business-as-usual" scenario of fossil fuel use.

### 2.2 $CO_2$ emissions scenarios

We illustrate five $CO_2$ emissions scenarios for the period 1850–2100. The first case, Business-As-Usual (BAU), assumes continuation of the ~2% annual growth of fossil fuel $CO_2$ emissions that has occurred in recent decades (*EIA*, 2006; *Marland et al.*, 2006). This 2% annual growth is assumed to continue for each of the three conventional fuels until ~half of each total reservoir (historic + remaining) has been exploited, after which emissions are assumed to decline 2% annually.

The second scenario, labeled Coal Phase-out, is meant to approximate a situation in which developed countries freeze their $CO_2$ emissions from coal by 2012 and a decade later developing countries similarly halt increases in coal emissions. Between 2025 and 2050 it is assumed that both developed and developing countries will linearly phase out emissions of $CO_2$ from coal usage. Thus in Coal Phase-out we have global $CO_2$ emissions from coal increasing 2% per year until 2012, 1% per year growth of coal emissions between 2013 and 2022, flat coal emissions for 2023–2025, and finally a linear decrease to zero $CO_2$ emissions from coal in 2050. These rates refer to emissions to the atmosphere and do not constrain consumption of coal, provided the $CO_2$ is captured and sequestered. Oil and gas emissions are assumed to be the same as in the BAU scenario.

The third, fourth, and fifth scenarios include the same phase-out of coal but investigate the effect of uncertainties in global oil usage and supply. The Fast Oil Use scenario adopts an alternative approach for calculating the peak in global oil emissions/usage, following the method of *Wood et al.* [2003]. It assumes that 2% annual growth in oil use continues past the midpoint of oil supplies, until the ratio of remaining reserves to emissions decreases to 10 yr from the current value of ~50 yr. This scenario causes 'peak oil' to be delayed ~21 years to 2037. The fourth scenario, Less Oil Reserves, uses the same trends as in Coal Phase-out but omits the oil 'reserve growth' term. This fourth scenario may be most relevant to a situation in which oil companies have been too optimistic about reserves and/or a high price on carbon emissions discourages exploration for oil in remote locations. Finally, the fifth scenario, Peak Oil Plateau, assumes that oil emissions exhibit a sustained peak from 2020–2040, using supply and $21^{st}$-century usage estimates of R. Nehring (*Kerr*, 2007). It assumes the global oil reserve base is ~50 Gt C larger than in our other scenarios (R. Nehring, personal communication, 14 June 2007; http://www.aapg.org/explorer/2007/05may/nehring.cfm). This scenario reflects the possibility that the peak productivity of major oil fields may occur at different times over the next several decades, leading to an extended, rather than abrupt, global oil peak.

In addition to fossil fuel $CO_2$ emissions, we also include historical and projected estimates for net emissions from land use in each of the above scenarios. We use historical (1850–2000) land use emissions estimates from CDIAC (*Houghton and Hackler*, 2002), and projections from the midrange IPCC Special Report on Emissions (SRES) A1T scenario (values from *IPCC*, 2001b).

These illustrative scenarios cover a broad range of fossil fuel reserves, but more extreme estimates do exist. We quantitatively investigate the effect of very low estimates of natural gas and crude oil reserves (*IPCC*, 2001a) and coal reserves (*WEC*, 2007). Although some analysts estimate reserves exceeding those in the illustrated scenarios, we do not study those because the cases we consider already reach far into the range of 'dangerous' atmospheric $CO_2$ levels.

## 2.3 Atmospheric $CO_2$ projections

For each of the $CO_2$ emissions scenarios we generate a time series of atmospheric $CO_2$ using the following parameterization of the Bern carbon cycle model of *Joos et al.* [1996]:

$$CO_2(t) = 18 + 14 \exp(-t/420) + 18 \exp(-t/70) + 24 \exp(-t/21) + 26 \exp(-t/3.4) \quad (1)$$

where $CO_2(t)$ is the percentage of emitted $CO_2$ remaining in the atmosphere after $t$ years, and the coefficients of each term are rounded from those provided in *Shine et al.* (2005). Note that Equation (1) implies that about one-third of anthropogenic $CO_2$ emissions remain in the atmosphere after 100 years and one-fifth after 1000 years.

Equation (1) is a static-sink PRF for anthropogenic $CO_2$ emissions, i.e. it is an approximation for the proportion (percent) of $CO_2$ remaining airborne $t$ years following an emission pulse. The time evolution of atmospheric $CO_2$ is obtained by taking the 1850 $CO_2$ concentration as 285.2 ppm (*Etheridge et al.*, 1998), recursively applying Equation (1) to the emission scenario, and integrating the results from 1850 to year $t$.

We also investigate the same scenarios using a dynamic-sink PRF that incorporates some carbon cycle feedbacks. Both approximations are based on the Bern carbon cycle model with the HILDA and 4-box biosphere models (equations 3-6, 16-17, A.2.2, and A.3 of *Joos et al.*, 1996). The dynamic-sink PRF includes simplified non-linear ocean carbonate chemistry as well as biospheric carbon uptake and respiration (both for a fixed climate). We demonstrate that the dynamic-sink PRF yields slightly different quantitative results than the static-sink PRF (eq. 1), but the differences are small enough to only reinforce our conclusions.

## 3. Results
### 3.1 Historical $CO_2$ emissions and concentrations

Integration of the product of Equation (1) and fossil fuel emissions over the period 1850-2007 yields an airborne fossil fuel $CO_2$ amount of ~85 ppm in 2007. The land use net emissions estimates of *Houghton and Hackler* [2002], with SRES A1T estimates for 2001–2007, yield an additional contribution of ~35 ppm in 2007. The $CO_2$ amount in 2007 is thus overestimated by ~20 ppm using eq. (1), or by ~13 ppm when the dynamic-sink PRF is used (Fig. 2). We suggest that, rather than a model deficiency, this overestimate is probably due to overestimate of net land use emissions (particularly for 1950–2000), based on the large uncertainties inherent in those estimates (*Houghton*, 2003), in contrast with the relatively high certainty of fossil fuel emissions estimates. Part of this discrepancy also may be due to the carbon cycle model. When the land use emissions are reduced by 50%, as supported by other studies (e.g., see Ch. 7 in *IPCC*, 2007), the model−data differences amount to at most ~2% from 1850−2007 (Fig. 2). Although part of the discrepancy could be a result of 'fertilization' of the biosphere, via anthropogenic $CO_2$ and nitrogen emissions, we show in a paper in preparation that incorporation of these effects in the carbon cycle model, as an alternative to reducing the land use source, has negligible impact on our present investigation. All of our scenarios therefore assume the above reduction in the *Houghton and Hackler* [2002] estimates.

This calculation for 1850-2007 provides a check on the reasonableness of the carbon cycle approximation (eq. 1) for $CO_2$ in the range 280-385 ppm. We infer that the model may continue to provide useful estimates for scenarios with moderate fossil fuel emissions, i.e., for the scenarios of special interest that keep atmospheric $CO_2$ less than or approximately 450 ppm.

As mentioned above and discussed in Section 3.4, for the larger $CO_2$ emissions of BAU scenarios, Equation (1) may begin to underestimate airborne $CO_2$, as it excludes the nonlinearity of the ocean carbon cycle as well as anticipated climate feedbacks on atmospheric $CO_2$ and $CH_4$, the latter being eventually oxidized to $CO_2$.

### 3.2 Projected $CO_2$ emissions

Table 1 provides an overview of the fossil fuel emissions and resulting atmospheric $CO_2$ amounts in our five emissions scenarios. We list peak fossil fuel emission years as ranges where necessary, to reflect minor differences in historical emissions estimates. (For example, there are differences between the historic emissions of CDIAC [*Marland et al.*, 2006] and *EIA* [2006]. Also, relatively minor changes arise from year to year in the CDIAC fossil fuel data due to retroactive updates to the UNSTAT database [T. Boden, personal communication, 5 June 2007].) Additional features of our four mitigation scenarios are listed in Table 2. Figure 3 shows $CO_2$ emissions over time for the five main scenarios and Figure 4 shows the resulting atmospheric $CO_2$ concentrations.

Peak oil emission in the BAU scenario occurs in 2016 ± 2 yr, peak gas in 2026 ± 2 yr, and peak coal in 2077 ± 2 yr (Fig. 3a). Coal Phase-out moves peak coal up to 2022 (Fig. 3b). Fast Oil Use causes peak oil to be delayed until 2037 (*Wood et al.*, 2003), but oil use then crashes rapidly (Fig. 3c). Less Oil Reserves results in peak oil moving to 2010 ± 2 yr (Fig. 3d), under the assumption that usage approximates the near symmetrical shape of the classical Hubbert curve. In the Peak Oil Plateau case, oil emissions peak in 2020 and remain at that level until 2040 (*Kerr*, 2007), thereafter decreasing approximately linearly (Fig. 3e).

Total fossil fuel $CO_2$ emissions peak in 2077 ± 2 yr at ~14 Gt C yr$^{-1}$ in the BAU scenario, almost double the current level, decreasing to ~9 Gt C yr$^{-1}$ in 2100 and ~3.3 Gt C yr$^{-1}$ in 2150 (figure 3(a)). Fuel emissions peak in 2016 ± 2 yr at ~10 Gt C yr$^{-1}$ in the Coal Phase-out scenario, decreasing to ~1 Gt C yr$^{-1}$ by 2100 (Fig. 3b). Cumulative 21$^{st}$-century fossil fuel emissions are ~1100 Gt C in the BAU scenario, ~500 Gt C in the Coal Phase-out scenario, ~520 Gt C in the Fast Oil use scenario, ~430 Gt C in the Less Oil Reserves scenario, and ~550 Gt C in the Peak Oil Plateau scenario.

### 3.3 Projected atmospheric $CO_2$ concentrations

Figure 4 shows atmospheric $CO_2$ concentrations resulting from our fossil fuel emissions scenarios. All of our scenarios include land use $CO_2$ emissions from the midrange SRES marker scenario (A1T). Additional calculations (not shown here) reveal that the land use emissions projections of the high-end scenario (A2) would add ~5 ppm to all of our $CO_2$ projections, while the low-end projections (B1) would cause a reduction by about the same amount.

Peak $CO_2$ in the BAU scenario is ~575 ppm in 2100, with fuel emissions alone raising $CO_2$ to over 560 ppm (Fig. 4a). This is more than double the pre-industrial $CO_2$ amount of ~280 ppm and already far past the 450 ppm threshold under consideration. Likely nonlinearities in the carbon cycle with such large $CO_2$ amounts would make the real-world peak $CO_2$ even greater, as would any contribution from unconventional fossil fuels.

Our interest is primarily in scenarios that limit atmospheric $CO_2$ to ~450 ppm or less. Therefore all scenarios other than BAU include phase-out of coal emissions at least as rapidly as in the standard Coal Phase-out scenario, which has moderate continued growth of global coal emissions until 2025, followed by linear phase-out of global emissions from coal between 2025 and 2050. This standard Coal Phase-out scenario has peak atmospheric $CO_2$ at ~445 ppm in 2046; fossil fuels alone raise $CO_2$ to ~428 ppm in 2047 (Fig. 4b).

The Fast Oil Use scenario (Fig. 4c) yields a peak atmospheric $CO_2$ level of ~463 ppm with fossil fuels raising $CO_2$ to ~446 ppm, i.e., faster use of the same oil amount increases the peak atmospheric amount by about 18 ppm. However, in the absence of carbon feedbacks, this difference decreases with time, practically disappearing by 2100.

The Less Oil Reserves scenario (Fig. 4d) yields a peak atmospheric $CO_2$ level of ~439 ppm, with fossil fuels raising $CO_2$ to ~422 ppm. Thus omission of oil reserve growth (Fig. 1) reduces the peak atmospheric $CO_2$ amount by ~6 ppm from the baseline Coal Phase-out scenario.

Lastly, the Peak Oil Plateau scenario (Fig. 4e) yields a peak atmospheric $CO_2$ level of ~456 ppm, with fossil fuels raising $CO_2$ to ~440 ppm. The sustained 20-yr peak in global oil emissions leads to an increase of ~10 ppm from the baseline Coal Phase-out case.

### 3.4 Additional scenarios considered

Effects of using the dynamic-sink PRF of *Joos et al.* [1996] are shown in Figures 5 and 6. The mean 1960–2007 total $CO_2$ airborne fraction of ~50% implied by the dynamic-sink PRF is lower than the ~53% based on the static-sink PRF, Eq. (1), and both are higher than the ~48% from observed airborne $CO_2$ and assumed fossil fuel and land use emissions (Fig. 5). The difference in the responses of the two PRFs to a pulse of 5 ppm $CO_2$ (~10 Gt C) is not major, although it does persist for centuries (Fig. 6a). Despite this, there is little difference (less than ~3%) between results from the two functions for the entire 21$^{st}$ century, regardless of whether high or moderate emissions are assumed (Fig. 6b,c). However, the BAU scenario with unrestrained emissions diminishes the buffering capacity of the ocean, leading to a ~20 ppm increase in peak $CO_2$ (Fig. 6b). On the other hand, with the lower emissions of the Coal Phase-out scenario, the oceanic/biospheric $CO_2$ uptake is similar for the two response functions (Fig. 6c).

We also computed time series for emissions and atmospheric $CO_2$ levels for several alternative sets of conventional oil, gas, and coal reserve estimates, including one from the World Energy Council (*WEC*, 2007) and one from *IPCC* (2001a, Table 3.28b). These estimates are lower than the *EIA* [2006] estimates (Fig. 1), and therefore, even assuming BAU growth and decline (Section 2.2), they yield earlier emissions peaks and lower peak atmospheric $CO_2$ levels (Fig. 7). Specifically, *WEC* (2007) coal reserves (~450 Gt C) yield peak coal emissions in ~2040, and IPCC (2001a) oil and gas reserves (118 Gt C and 82 Gt C, respectively) yield peak oil emissions in ~2004 and peak gas emissions in ~2009 (Fig. 7a).

Assuming the *IPCC* [2001a] oil and gas reserves along with *WEC* [2007] coal reserves yields a peak atmospheric $CO_2$ level of ~457 ppm in 2076, with fossil fuels alone raising $CO_2$ to ~442 ppm (Fig. 7b). Although this latter scenario represents relatively moderate 'BAU' cases, it relies heavily on the assumption that carbon-positive substitute fuels cannot or will not be developed in the future to replace declining conventional fuel reserves, e.g., due to a rising price on carbon emissions (see Discussion).

### 3.5 Comparison with IPCC-SRES and EMF-21 scenarios

In contrast with all of the above scenarios, peak total emissions in the four SRES scenario families (*IPCC*, 2000) range from ~12 Gt C $yr^{-1}$ in 2040 (B1 marker scenario) to a staggering ~28 and 29 Gt C $yr^{-1}$ in 2100 (A2 and A1F1 marker scenarios, respectively). Time-integrated $21^{st}$-century emissions for these SRES marker scenarios range from ~970 Gt C (B1) to ~1900 and 2100 Gt C (A2 and A1F1). Thus, it is clear that the high-end SRES scenarios implicitly assume that, in the absence of climate mitigation policies, massive amounts of unconventional or 'undiscovered' resources will become viable substitutes for dwindling conventional reserves.

Resulting atmospheric $CO_2$ amounts in 2100 in the SRES scenarios range from ~540 ppm to ~970 ppm, excluding carbon cycle feedbacks (*IPCC*, 2001b). Model simulations suggest that carbon cycle feedbacks under a high-end emissions scenario (A2) can yield an additional 20-200 ppm of $CO_2$ by 2100 (*Friedlingstein et al.*, 2006). Note that our four mitigation scenarios, however, are consistent with current assessments of the cumulative $21^{st}$-century emissions needed to stabilize atmospheric $CO_2$ at 450 ppm even after factoring in carbon cycle feedbacks (e.g., see *IPCC*, 2007).

Our scenarios also differ from those of EMF-21 (*Weyant et al.*, 2006) in several ways. First, their scenarios have a long-term global radiative forcing target of ~4.5 W $m^{-2}$ above the pre-industrial level, which yields a target global temperature rise of over 3°C, for nominal estimates of climate sensitivity (*Hansen et al.*, 2007b; *IPCC*, 2007). Also, their mean model ensemble 21st-century $CO_2$ emissions are assumed to continually increase from ~14 Gt C $yr^{-1}$ in 2025 to ~25 Gt C $yr^{-1}$ by 2100. Thus, like the SRES scenarios, the EMF-21 scenarios reflect climate change that we would classify as 'dangerous' and therefore highly undesirable.

## 4. Discussion
### 4.1 Avoidance of 'dangerous' anthropogenic climate change

Practically all nations of the world have agreed that a 'dangerous' increase of atmospheric greenhouse gases should be avoided (*UNFCCC*, 1992), but the dangerous level of gases is not well defined. *Hansen et al.* [2007a,b] have argued that additional global warming above that in 2000 must be kept less than 1°C, and that, therefore, the dangerous $CO_2$ level is at most about 450 ppm, and likely less than that. Although moderate trade-off with non-$CO_2$ gases is possible (*Hansen and Sato*, 2004), $CO_2$ is the most important climate forcing because a considerable fraction of fossil fuel $CO_2$ emissions remains in the atmosphere for many centuries (*Archer*, 2005). Given that $CO_2$ has already increased during the industrial era from ~280 ppm to ~385 ppm, there is some urgency in determining what steps are practical to limit further growth of atmospheric $CO_2$. Indeed, the scenarios used in climate projections by the *IPCC* [2000, 2001a,b, 2007] all have $CO_2$ increasing well beyond 450 ppm.

On the other hand, fossil fuel reservoirs are finite, and existing deposits do not have to be fully exploited. Also, it may be practical to capture and sequester much of the $CO_2$ emitted in burning coal at power plants. Thus it is important to estimate expected atmospheric $CO_2$ levels for realistic estimates of fossil fuel reserves and to determine how the $CO_2$ level depends upon possible constraints on coal use.

We suggest that, for the sake of simplicity and transparency, it is useful to make such estimates with simple pulse response functions for airborne $CO_2$, although similar studies should also be made with comprehensive carbon cycle models. We view the pulse response functions that we have employed as providing an approximate lower bound for the proportion of fossil fuel $CO_2$ emissions that remain airborne. The uptake capacity of the carbon sinks may decrease if the $CO_2$ source increases, and there are potential climate feedbacks that could add $CO_2$ to the atmosphere, e.g. carbon emissions from forest dieback (*Cox et al.*, 2000), melting permafrost (*Walter et al.*, 2006; *Zimov et al.*, 2006), and warming ocean floor (*Archer*, 2007). In addition, land use change and deforestation may remain a significant source of positive anthropogenic climate forcing this century, e.g., *Gruber et al.* [2004] assert that land use-related human activities could lead to the release of up to ~40 Pg C over the next ~20 years and ~100 Pg C this century due to the alteration of live biomass pools in tropical and subtropical ecosystems.

On the other hand, if fossil fuel emissions of $CO_2$ decrease, concerns about possible non-linear positive feedbacks are diminished. Indeed, we suggest a possible dichotomy of scenarios: if $CO_2$ emissions decrease, the proportion of $CO_2$ taken up by sinks could increase, with a resulting climate forcing that is much less than that in scenarios with continually increasing $CO_2$ emissions.

Given these basic considerations, we have focused on scenarios in which coal use is phased out except where the $CO_2$ is captured. We find that, with such an assumption, it is possible to keep maximum 21$^{st}$-century atmospheric $CO_2$ less than 450 ppm, provided that the EIA estimates of oil and gas reserves and reserve growth are not significant underestimates. This limit on $CO_2$ is achieved in our scenarios only if cumulative global emissions from coal between the present and 2050 amount to ~100 Gt C or less. Thus, even if coal reserves are much lower than historically assumed (e.g., *NRC*, 2007), there is surely enough coal to take the world past 450 ppm $CO_2$ without mitigation efforts such as those described here. On the bright side, our findings indicate that a feasible time scale for reductions can keep $CO_2$ below 450 ppm.

### 4.2 Future use of fossil fuels

Goals for atmospheric $CO_2$ amount surely must be adjusted as knowledge about climate change and its impacts improves. Recent evidence of sea ice loss in the Arctic and accelerating net mass loss from the West Antarctic and Greenland ice sheets suggest that the allowable level of warming is likely less than 1°C above the 2000 global temperature and the $CO_2$ limit is likely less than 450 ppm. Thus, details about the magnitude of fossil fuel reserves and the rate at which the reserves are exploited may be important. We find that the maximum $21^{st}$-century atmospheric $CO_2$ level varies by ~18 ppm depending upon the rate at which given oil and gas reserves are consumed. This variation decreases with time, however, so the size of the exploited oil and gas reservoirs is a more important consideration.

The size of economically recoverable oil and gas resources is flexible, depending upon the degree to which fossil fuels are priced to cover their environmental costs. Thus we have argued (*Hansen et al.*, 2007a; *Hansen*, 2007) in favor of placing a significant rising price on $CO_2$ emissions. One effect of a rising carbon price would be to slow the rate at which fossil fuel resources are exploited, thus reducing the maximum atmospheric $CO_2$ amount, as illustrated above. More importantly, a carbon price would result in some of the oil, gas, and coal being left in the ground, primarily deposits at great depths or in extreme environmental locations. Given that the world must move beyond fossil fuels for its energy needs eventually, it is appropriate to encourage that transition soon, and thus minimize anthropogenic climate change. We note that there are various ways of placing a price on carbon emissions, such as a progressive carbon tax, industry 'cap and trade' measures, or individual 'ration and trade' measures. The pricing scheme should be chosen based on economic effectiveness and fairness.

*Hirsch et al.* [2005] note that it requires decades to remake energy infrastructure, and thus peaking of oil and gas production has the potential for severe economic disruption if steps are not taken to encourage technology development and implementation. This consideration adds to the need for prompt actions to conserve readily available oil and gas, thus stretching out these conventional supplies, while encouraging innovations in energy efficiency and alternative (non-fossil) energies. Stretching of supplies is a principal function of an increasing carbon price. Nuclear power could be one viable alternative option, if strict provisions are followed for public safety, waste disposal, and elimination of potential weapons-grade by-products; adoption of an international nuclear environmental treaty could be a significant step toward this end (*Robock et al.*, 2007).

**4.3 Additional climate change mitigation measures**

Finally, we note that, as understanding of climate change and its impacts improves, it is possible that even lower limits on atmospheric $CO_2$ and the net anthropogenic climate forcing than discussed here may prove to be highly desirable. It has been suggested that to buy extra time to enact such large-scale mitigation, societies should adopt an approach that incorporates both emissions reductions and geoengineering options like periodic, sustained stratospheric sulfate aerosol injection (*Crutzen*, 2006; *Wigley*, 2006). However, a geoengineering "quick fix", if not sustained precisely to the degree and length of time needed, could do more harm than good (*Matthews and Caldeira*, 2007), and it is difficult to define how much "fix" is needed. Thus, while geoengineering might provide some benefit, the potential gain must be weighed against long-term risks to climate and oceanic/stratospheric chemistry. Especially given the existence of low-cost and no-cost methods to reduce $CO_2$ emissions (*Lovins*, 2005), slowing of fossil fuel $CO_2$ emissions warrants highest priority.

Further reductions of anthropogenic climate forcing, beyond the 2025–2050 coal phase-out strategy that we quantified here, could be achieved as follows:

(1) A freeze on new construction of traditional coal-fired power plants (without $CO_2$ sequestration) by 2010, with a linear phase-out of all such existing plants between 2010 and 2030. This action reduces the maximum atmospheric $CO_2$ from ~445 ppm in our standard Coal

Phase-out scenario to ~400-430 ppm (depending on oil and gas reserve size). Fossil fuel contribution to atmospheric $CO_2$ level decreases from ~428 ppm to ~390-410 ppm.

(2) Intensive efforts to reduce non-$CO_2$ anthropogenic climate forcings, especially methane, tropospheric ozone, and black carbon. *Hansen and Sato* [2004] estimate that realistic potential savings from such reductions are equivalent to 25−50 ppm of $CO_2$.

(3) Anthropogenic draw-down of atmospheric $CO_2$. Farming and forestry practices that enhance carbon retention and storage in the soil and biosphere should be supported (*McCarl and Sands*, 2007), as should large-scale reforestation. Direct removal of $CO_2$ from the air through expedited carbonate formation also holds great potential (*Lackner*, 2003; *Keith et al.*, 2006). In addition, burning biofuels in power plants with carbon capture and sequestration can draw down atmospheric $CO_2$ (*Hansen*, 2007), in effect putting anthropogenic $CO_2$ back underground where it came from. However, careful measures must be taken to ensure that biofuel production does not occur at the expense of food crops and tropical forests are not converted to biofuel farms. For instance, agricultural waste, natural grasses and other cellulosic material can be used (e.g., *Tilman et al.*, 2006). Fertilizers used in their production should minimize emission of non-$CO_2$ greenhouse gases as well. $CO_2$ sequestered beneath ocean sediments is inherently stable (*House et al.*, 2006), and other safe geologic sites are also available.

**Acknowledgements**

We thank Makiko Sato for providing the extrapolated historic $CO_2$ emissions data for 2004 and 2005 as well as the compiled historical $CO_2$ concentration data. We also thank Gioietta Petravic, Fortunat Joos, Jackson Harper, Dave Rutledge, and two anonymous reviewers for helpful comments on the manuscript. Research support was provided by Hal Harvey of the Hewlett Foundation, Gerry Lenfest, and NASA Earth Science Research Division managers Jack Kaye and Don Anderson.

**References**

Archer, D. (2005), Fate of fossil fuel $CO_2$ in geologic time, *J. Geophys. Res.*, *110*, C09 S05, doi:10.1029/2004JC002625

Archer, D. (2007), Methane hydrate stability and anthropogenic climate change, *Biogeosc. Discuss.*, *4*, 993-1057.

British Petroleum (BP) (2006), Putting energy in the spotlight: BP Statistical Review of World Energy, June 2006, www.bp.com/pdf/statistical_review_of _world energy_full_report2006.pdf.

Conway, T.J., P.M. Lang, and K.A. Masarie (2007), Atmospheric Carbon Dioxide Dry Air Mole Fractions from the NOAA ESRL Carbon Cycle Cooperative Global Air Sampling Network, 1968-2006, Version: 2007-09-19, Path: ftp://ftp.cmdl.noaa.gov/ccg/co2/flask/event/

Crutzen, P. (2006), Albedo enhancement by stratospheric sulfur injections: A contribution to resolve a policy dilemma?, *Clim. Change*, *77*, 211-219.

Energy Information Administration (EIA), U.S. Dept. of Energy (2006), *International Energy Outlook 2006*, http://www.eia.doe.gov/oiaf/archive/ieo06/index.html

Etheridge, D.M., L.P. Steele, R.L. Langenfelds, R.J. Francey, J.-M. Barnola and V.I. Morgan (1998), Historical $CO_2$ records from the Law Dome DE08, DE08-2, and DSS ice cores, in *Trends: A Compendium of Data on Global Change*, Carbon Dioxide Information Analysis Center, Oak Ridge National Laboratory, U.S. Department of Energy, Oak Ridge, Tenn., U.S.A.

Friedlingstein, P., et al. (2006), Climate-carbon cycle feedback analysis: Results from the $C^4MIP$ model intercomparison, *J. Climate*, *19*, 3337–3353.

Gruber, N., P. Friedlingstein, C. B. Field, R. Valentini, M. Heimann, J. E. Richey, P. R. Lankao, E-D. Schulze, and C. Chen (2004), The vulnerability of the carbon cycle in the 21st century: an assessment of carbon-climate-human interactions, in *The Global Carbon Cycle*, edited by C. B. Field and M. R. Raupach, Island Press, Washington, pp. 45-76.

Hansen, J. E. (2007), Dangerous human-made interference with climate, Testimony to Select Committee on Energy Independence and Global Warming, United States House of Representatives, 26 April 2007 (available at http://globalwarming.house.gov/list/hearing/global_warming/hearing_070423.shtml; updated version available at http://arxiv.org/abs/0706.3720).

Hansen, J. E., and M. Sato (2004), Greenhouse gas growth rates, *Proc. Natl. Acad. Sci.*, *101*, 16,109-16,114.

Hansen, J. E., et al. (2007a), Dangerous human-made interference with climate: a GISS modelE study, *Atmos. Chem. Phys.*, *7*, 2287-2312.

Hansen, J. E., M. Sato, P. Kharecha, G. Russell, D. W. Lea, and M. Siddall (2007b), Trace gases and climate change, *Phil. Trans. Royal Soc. A*, *365*, 1925-1954, doi:10.1098/rsta.2007.2052

Hirsch, R.L., R. Bezdek, R. Wendling (2005), Peaking of world oil production: impacts, mitigation, and risk management, report to U.S. Dept. of Energy – Natl. Energy Technol. Lab. (avail. at http://www.netl.doe.gov/publications/others/pdf/Oil_Peaking_NETL.pdf ).

Houghton, R.A., and J.L. Hackler (2002), Carbon Flux to the Atmosphere from Land-Use Changes, in *Trends: A Compendium of Data on Global Change*, Carbon Dioxide Information Analysis Center, Oak Ridge National Laboratory, U.S. Department of Energy, Oak Ridge, Tenn., U.S.A.

Houghton, R. A. (2003), Revised estimates of the annual net flux of carbon to the atmosphere from changes in land use and land management 1850–2000, *Tellus*, 55*B*, 378-390.

House, K.Z., D.P. Schrag, C.F. Harvey, and K.S. Lackner (2006), Permanent carbon dioxide storage in deep-sea sediments, *Proc. Natl. Acad. Sci.*, *103(33)*, 12291, doi:10.1073/pnas.0605318103

Hubbert, M.K. (1956), Nuclear energy and the fossil fuels, *Publication no. 95*, 40 pp., Shell Development Company, Houston, Tex.

Intergovernmental Panel on Climate Change (IPCC) (2000), *Special Report on Emissions Scenarios*, edited by N. Nakicenovic and R. Swart, Cambridge Univ. Press, Cambridge, U.K.

Intergovernmental Panel on Climate Change (IPCC) (2001a), *Climate Change 2001: Mitigation*, edited by B. Metz, O. Davidson, R. Swart, and J. Pan, Cambridge Univ. Press, Cambridge, U.K.

Intergovernmental Panel on Climate Change (IPCC) (2001b), *Climate Change 2001: The Scientific Basis*, edited by J.T. Houghton, Y. Ding, D.J. Griggs, M. Noguer, P.J. van der Linden, X. Dai, K. Maskell, and C.A. Johnson, Cambridge Univ. Press, Cambridge, U.K.

Intergovernmental Panel on Climate Change (IPCC) (2007), *Climate Change 2007: The Physical Science Basis. Contribution of Working Group I to the Fourth Assessment Report of the Intergovernmental Panel on Climate Change*, edited by S. Solomon, D. Qin, M. Manning, Z. Chen, M. Marquis, K.B. Averyt, M. Tignor and H.L. Miller, Cambridge Univ. Press, Cambridge, U.K.

Joos, F., M. Bruno, R. Fink, T.F. Stocker, U. Siegenthaler, C. Le Quere, and J.L. Sarmiento (1996), An efficient and accurate representation of complex oceanic and biospheric models of anthropogenic carbon uptake, *Tellus*, *48B*, 397-417.

Keeling, C.D. and T.P. Whorf. (2005), Atmospheric $CO_2$ records from sites in the SIO air sampling network, in *Trends: A Compendium of Data on Global Change*, Carbon Dioxide Information Analysis Center, Oak Ridge National Laboratory, U.S. Department of Energy, Oak Ridge, Tenn., U.S.A.

Keith, D.W., M.H. Duong, and J.K. Stolaroff (2006), Climate strategy with $CO_2$ capture from the air, *Clim. Change*, *74*, 17-45.

Kerr, R.A. (2005), Bumpy road ahead for world's oil, *Science*, *310*, 1106-1108.

Kerr, R.A. (2007), The looming oil crisis could arrive uncomfortably soon, *Science*, *316*, 351.

Lackner, K. (2003), A guide to $CO_2$ sequestration, S*cience*, *300*, 1677-1678.

Lam, M. (1998), Louisiana short term oil and gas forecast, report to Louisiana Dept. of Natural Resources – Technol. Assessment Div. http://dnr.louisiana.gov/sec/execdiv/techasmt/oil_gas/forecasts/shortterm_1998/03-production.htm.

Lovins, A. B. (2005), More profit with less carbon, *Scientific American*, *293*, 74-82.

Marland, G., T. A. Boden, and R. J. Andres (2006), Global, regional, and national $CO_2$ emissions, in *Trends: Compendium of Data of Global Change*, Carbon Dioxide Information Analysis Center, Oak Ridge National Laboratory, U.S. Dept. Energy, Oak Ridge, TN.

Marshall, A. (1890), *Principles of Economics*, Macmillan and Co., London.

Matthews, H. D., and K. Caldeira (2007), Transient climate-carbon simulations of planetary geoengineering, *Proc. Natl. Acad. Sci.*, *104*: 9949-9954 doi:10.1073/pnas.0700419104

McCarl, B. A. and R. D. Sands (2007), Competitiveness of terrestrial greenhouse gas offsets: are they a bridge to the future?, *Clim. Change*, *80*, 109-126.

Milici, R. C., and E. V. M. Campbell (1997), A predictive production rate life-cycle model for southwestern Virginia coalfields, *USGS Circular 1147* http://pubs.usgs.gov/circ/c1147/

National Research Council (NRC) (2007), *Coal: Research and Development to Support National Energy Policy*, The National Academies Press, Washington.

Robock, A., O. B. Toon, R. B. Turco, L. Oman, G. L. Stenchikov, and C. Bardeen (2007), The continuing threat of nuclear weapons: Integrated policy responses, *Eos*, *88*, 228.

Shine, K. P., J. S. Fuglestvedt, K. Hailemariam, and N. Stuber (2005), Alternatives to the global warming potential for comparing climate impacts of emissions of greenhouse gases, *Clim. Change*, *68*, 281-302.

Thoning, K.W., D.R. Kitzis, and A. Crotwell (2007), Atmospheric Carbon Dioxide Dry Air Mole Fractions from quasi-continuous measurements at Barrow, Alaska; Mauna Loa, Hawaii; American Samoa; and South Pole, 1973-2006, Version: 2007-10-01, Path: ftp://ftp.cmdl.noaa.gov/ccg/co2/in-situ/

Tilman, D., J. Hill, and C. Lehman (2006), Carbon-negative biofuels from low-input high-diversity grassland biomass, *Science*, *314*, 1598-1600.

United Nations Framework Convention on Climate Change (UNFCCC), 1992 http://unfccc.int/essential_background/convention/background/items/1349.php

van der Veen, C. J. (2006), Reevaluating Hubbert's prediction of U.S. peak oil, *Eos*, *87*, 199.

Walter, K. M., S. A. Zimov, J. P. Chanton, D. Verbyla, and F. S. Chapin (2006), Methane bubbling from Siberean thaw lakes as a positive feedback to climate, *Nature*, *443*, 71-75.

Weyant, J. P., F. C. de la Chesnaye, and G. J. Blanford (2006), Overview of EMF-21: Multigas mitigation and climate policy, *The Energy Journal*, Special Issue, 1-32.

Wigley, T. M. (2006), A combined mitigation/geoengineering approach to climate stabilization, *Science*, *314*, 452-454.

Wood, J. H., G. Long, and D. Morehouse (2003), World conventional oil supply expected to peak in 21st century, *Offshore*, *63*, 90 (online version available at http://www.eia.doe.gov/pub/oil_gas/petroleum/feature_articles/2004/worldoilsupply/oilsupply04.html)

World Energy Council (2007), *Survey of Energy Resources* (21st ed.), edited by J. Trinnaman and A. Clarke http://www.worldenergy.org/publications/survey_of_energy_resources_2007/default.asp

Zimov, S. A., E. A. G. Schuur, and F. S. Chapin (2006), Permafrost and the global carbon budget, *Science*, *312*, 1612-1613.

Table 1. Approximate peak fossil fuel $CO_2$ emissions and atmospheric $CO_2$ levels in each scenario.

| Scenario | Peak fuel emission | Year of peak | Peak fuel $CO_2$ level | Year of peak |
|---|---|---|---|---|
| **BAU** | 14 Gt C yr$^{-1}$ | 2077 ± 2 yr | 563 ppm | 2100 |
| **Coal Phase-out** | 10 Gt C yr$^{-1}$ | 2016 ± 2 yr | 428 ppm | 2047 |
| **Fast Oil Use** | 11 Gt C yr$^{-1}$ | 2025 ± 2 yr | 446 ppm | 2046 |
| **Less Oil Reserves** | 9 Gt C yr$^{-1}$ | 2022 ± 2 yr | 422 ppm | 2045 |
| **Peak Oil Plateau** | 10 Gt C yr$^{-1}$ | 2025 ± 2 yr | 440 ppm | 2060 |

Table 2. Salient features and metrics of mitigation scenarios.

| Scenario | 2007−2050 coal emissions | Total 2007−2050 fuel emissions | Reduction in 2050 vs. 2007 fuel emissions |
|---|---|---|---|
| **Coal Phase-out** | ~110 Gt C | ~330 Gt C | 57% |
| **Fast Oil Use** | ~110 Gt C | ~390 Gt C | 54% |
| **Less Oil Reserves** | ~110 Gt C | ~300 Gt C | 66% |
| **Peak Oil Plateau** | ~110 Gt C | ~360 Gt C | 40% |

**Figures**

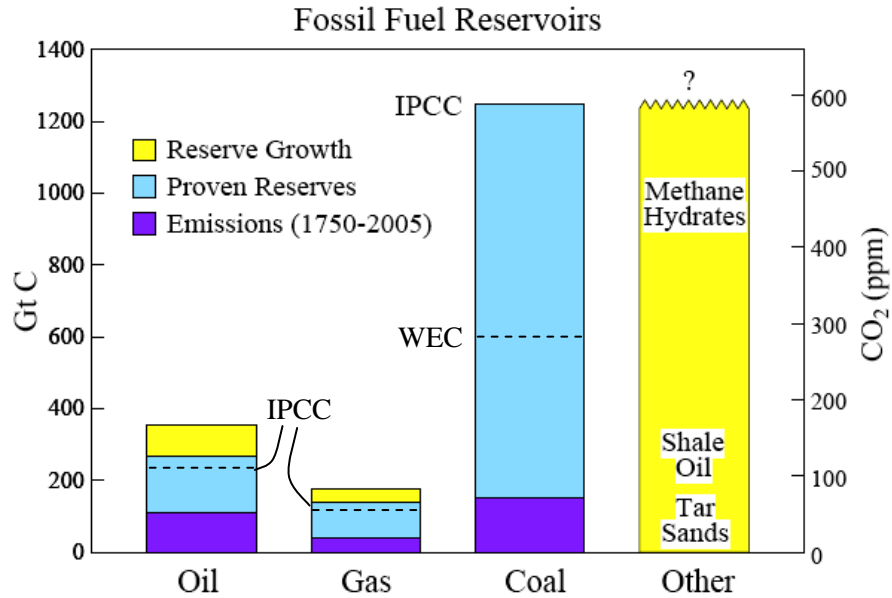

Figure 1. Fossil fuel-related estimates used in this study. Historical fossil fuel $CO_2$ emissions from the Carbon Dioxide Information Analysis Center (CDIAC; *Marland et al.*, 2006) and British Petroleum (*BP*, 2006). Lower limits for current proven conventional reserve estimates for oil and gas from *IPCC* [2001a] (dashed lines), upper limits and reserve growth values from US Energy Information Administration (*EIA*, 2006). Lower limit for conventional coal reserves from World Energy Council (*WEC*, 2007; dashed line), upper limit from *IPCC* [2001a]. Possible amounts of unconventional fossil resources from *IPCC* [2001a].

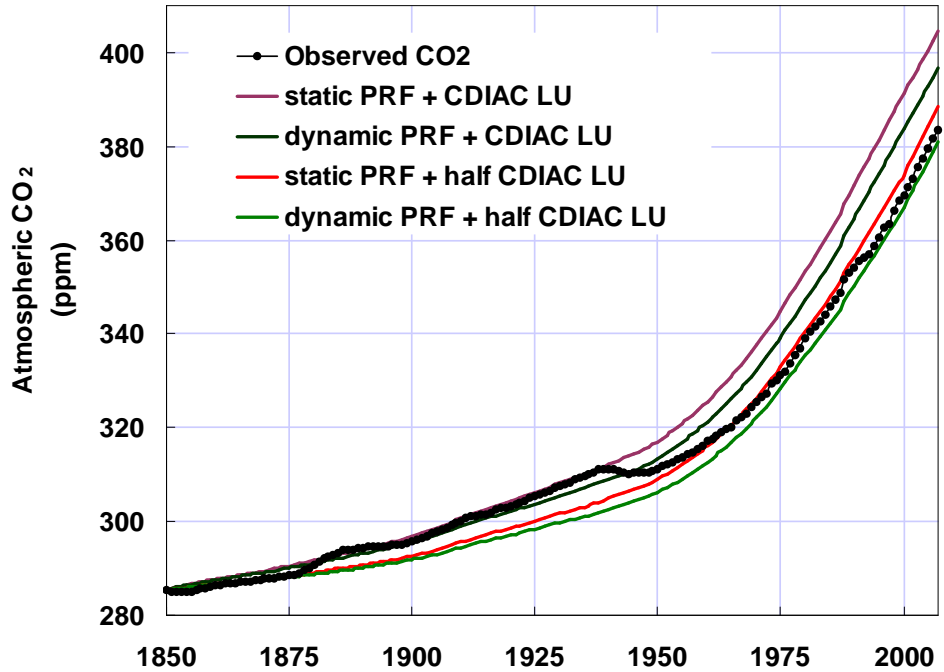

Figure 2. Computed vs. observed time evolution of industrial-era atmospheric $CO_2$ from 1850–2007. (CDIAC LU = *Houghton and Hackler* [2002] land use emissions for 1850–2000). Inclusion of the CDIAC LU emissions causes increasing overestimation of $pCO_2$ by the model between 1950–2000 for both the static-sink and dynamic-sink PRFs, suggesting that those LU estimates may be overestimates (see Section 3.1). When the CDIAC LU estimates are reduced by 50%, both PRFs produce very good agreement with observed $CO_2$. Observations prior to 1958 are based on Law Dome ice core data (*Etheridge et al.*, 1998), and from 1958 onwards based on high-precision flask and in-situ measurements (*Keeling and Whorf*, 2005; *Conway et al.*, 2007; *Thoning et al.*, 2007), with the specific data series as concatenated and adjusted to global means by *Hansen and Sato* [2004].

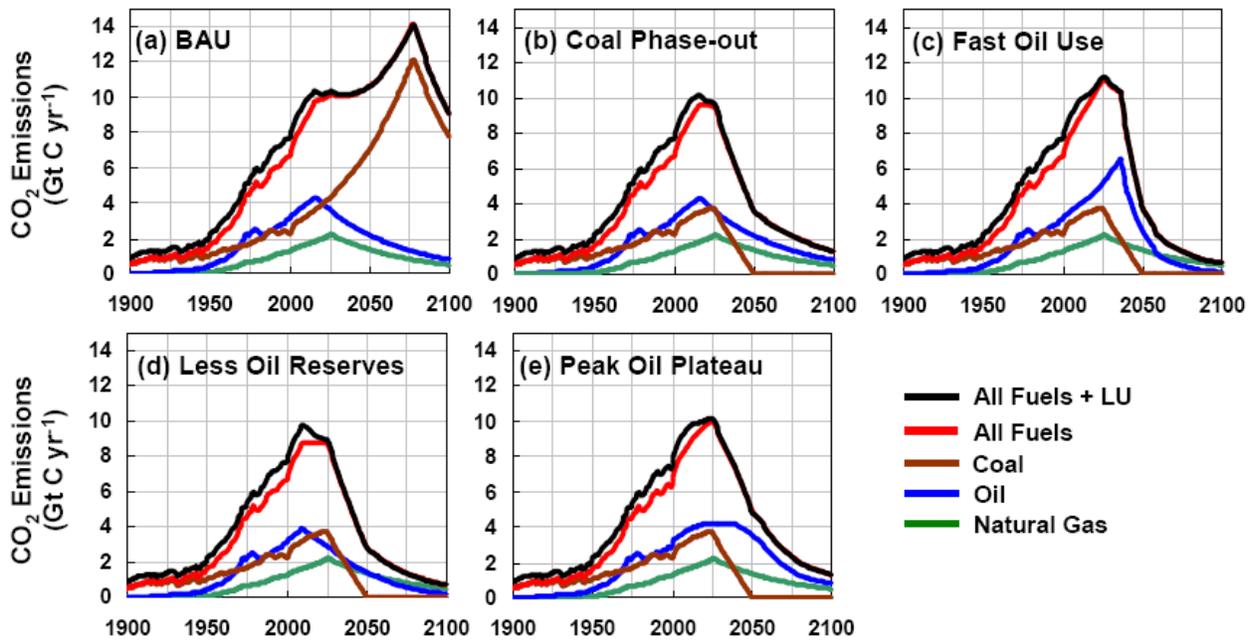

Figure 3. Historical and projected anthropogenic $CO_2$ emissions for the five main scenarios (LU = land use).

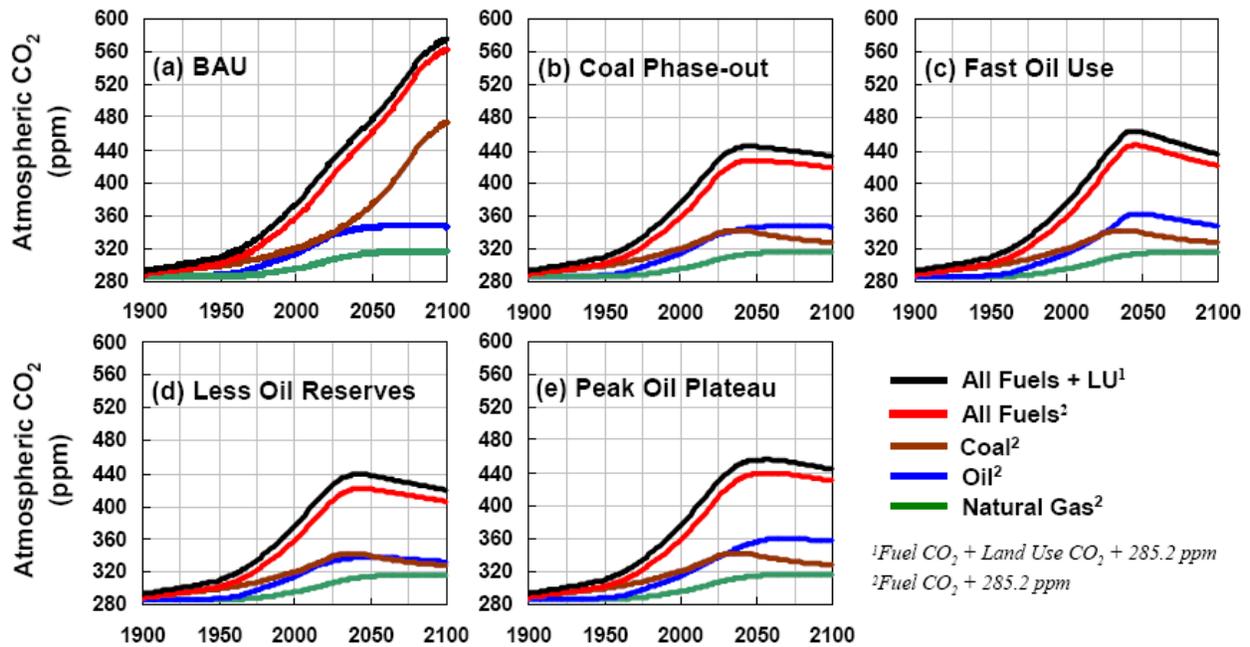

Figure 4. Time evolution of atmospheric $CO_2$ concentrations for the five main case scenarios. In each scenario, the computed amount of $CO_2$ contributed by each source is added to the baseline 1850 $CO_2$ level of 285.2 ppm to generate the individual curves. Compared with the control case (Coal Phase-out, panel (b)), the peak $CO_2$ level in the Less Oil Reserves scenario is ~6 ppm lower, while the peak level in the Fast Oil Use and Peak Oil Plateau cases is ~18 ppm and ~10 ppm higher, respectively.

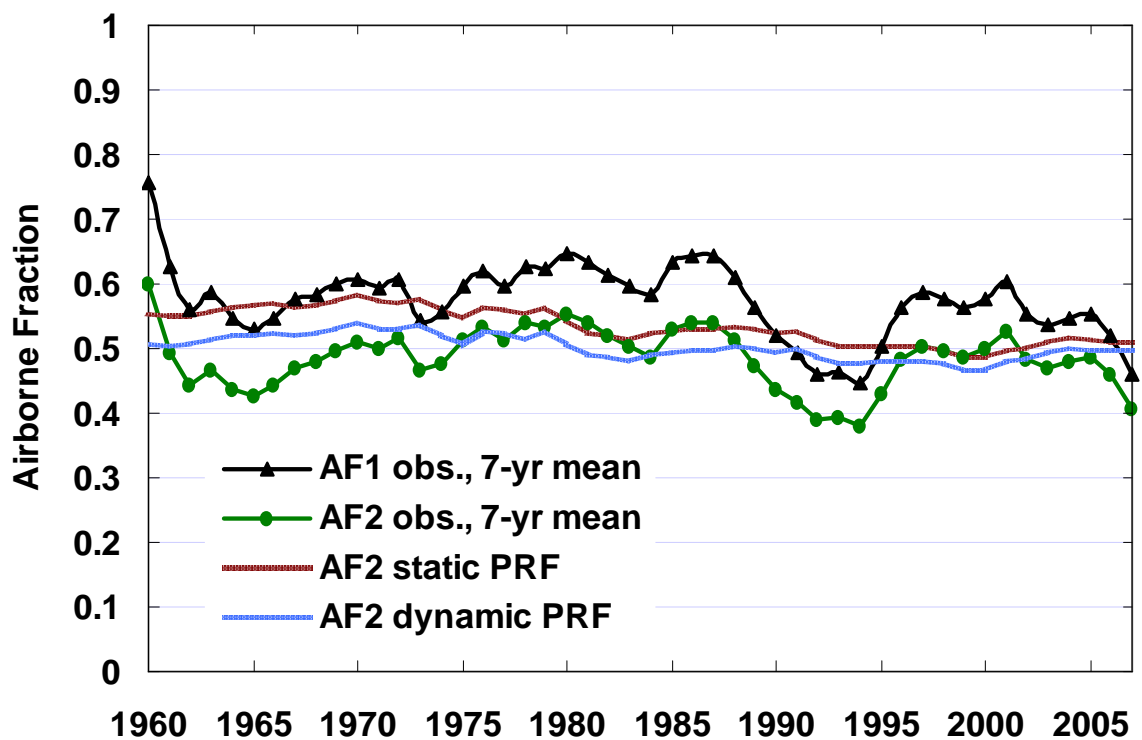

Figure 5. $CO_2$ 'airborne fraction' (AF) of 1960–2007 anthropogenic $CO_2$ emissions, computed as the mean measured atmospheric $CO_2$ concentration of a given year minus the amount in the previous year divided by either the fossil fuel emissions in the given year (AF1) or the sum of fossil fuel and land use emissions in the given year (AF2). The 1960–2007 mean derived from observed $CO_2$ is ~57% for the former (AF1 obs.) and ~48% for the latter (AF2 obs.). For the static-sink and dynamic-sink PRFs, the 1960–2007 model mean AF2 values are ~53% and ~50%, respectively. (See Fig. 2 caption for $CO_2$ data sources.)

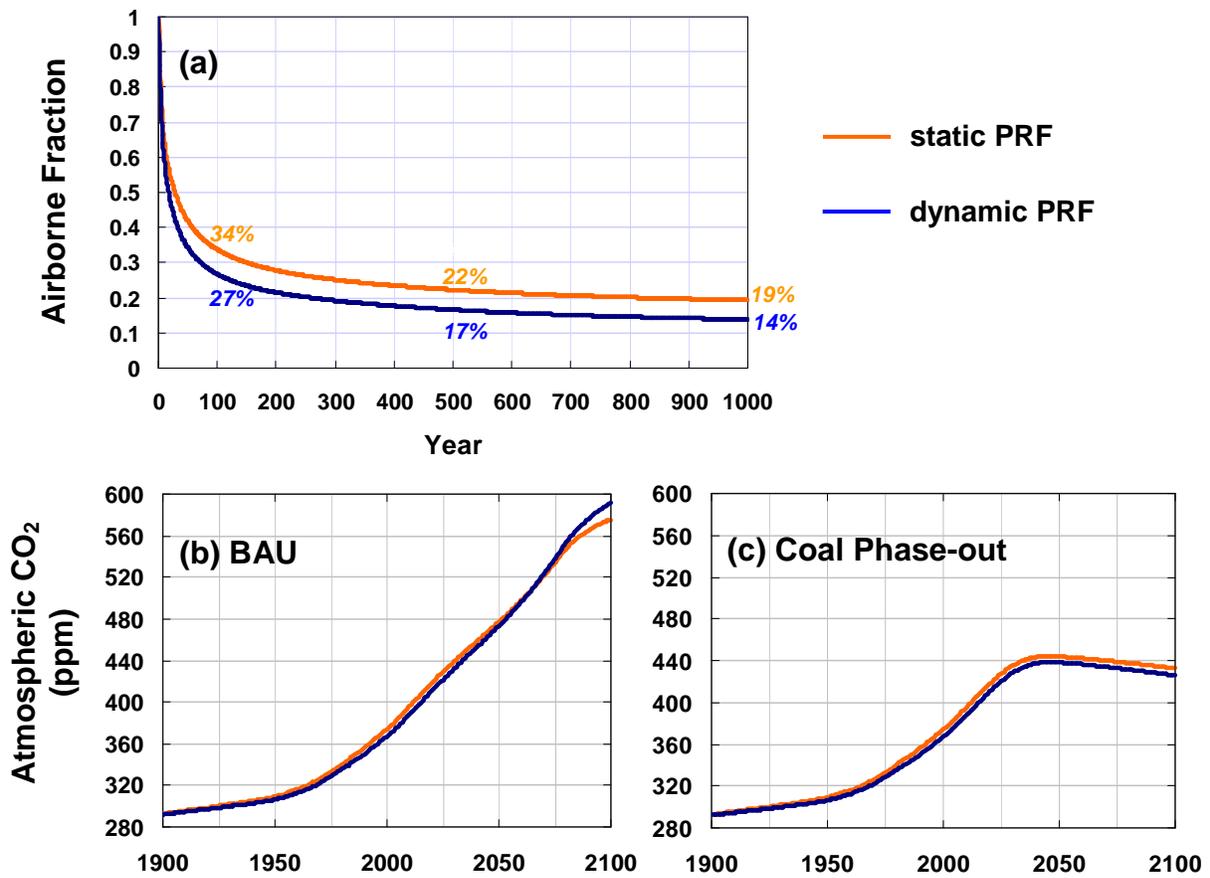

Figure 6. Effects of using the static-sink PRF vs. the dynamic-sink PRF. (a) Annual fraction of $CO_2$ remaining airborne after a pulse emission of 5 ppm (~10 Gt C), with differences highlighted at 100, 500, and 1000 yr. Atmospheric $CO_2$ concentrations for the baseline (b) BAU and (c) Coal Phase-out scenarios. The difference is generally negligible (less than ~3% throughout the 21st century), but in the high-emission BAU scenario the dynamic-sink PRF yields ~20 ppm greater peak $CO_2$ in 2100.

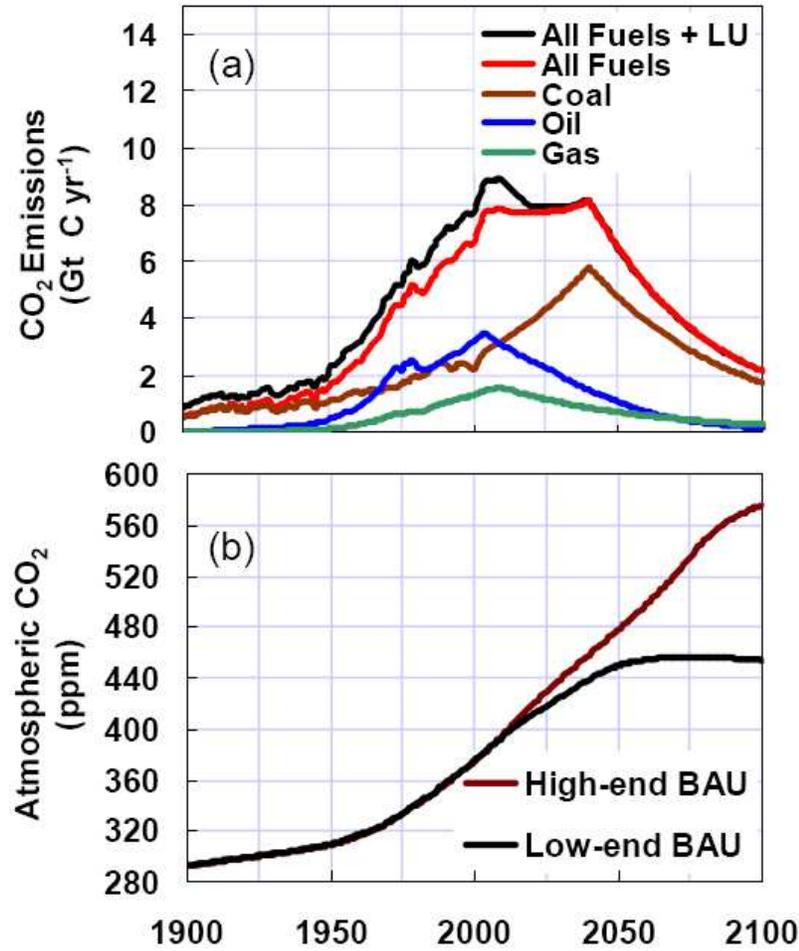

Figure 7. (a) Alternate 'low-end' BAU emissions scenario assuming conventional oil and gas reserves from *IPCC* (2001a) with no reserve growth, and coal reserves from World Energy Council [*WEC*, 2007] (LU=land use). (b) Resulting atmospheric $CO_2$ from this scenario, compared with the baseline ('high-end') BAU scenario from Fig. 4a.